\documentclass{article}
\usepackage[latin1]{inputenc}
\usepackage{dcolumn}
\usepackage{bm}
\usepackage{verbatim}       

\usepackage[dvips]{graphicx}
\usepackage{amsthm}
\usepackage{amssymb}
\usepackage{bm}
\usepackage{amstext}
\usepackage{amsmath}
\usepackage{amsfonts}
\usepackage{dsfont}
\usepackage{txfonts}
\usepackage{color}
\usepackage{mathrsfs}

\newfont{\bb}{msbm10 at 12pt}

\newcommand{\bd}{\begin{definition}}                
\newcommand{\ed}{\end{definition}}                  
\newcommand{\bc}{\begin{corollary}}                 
\newcommand{\ec}{\end{corollary}}                   
\newcommand{\bl}{\begin{lemma}}                     
\newcommand{\el}{\end{lemma}}                       
\newcommand{\bp}{\begin{proposition}}            
\newcommand{\ep}{\end{proposition}}                
\newcommand{\bere}{\begin{remark}}                  
\newcommand{\ere}{\end{remark}}                     

\newcommand{\bt}{\begin{theorem}}
\newcommand{\et}{\end{theorem}}

\newcommand{\be}{\begin{equation}}
\newcommand{\ee}{\end{equation}}

\newcommand{\bit}{\begin{itemize}}
\newcommand{\eit}{\end{itemize}}
\newtheorem{theorem}{Theorem}[section]
\newtheorem{corollary}[theorem]{Corollary}

\newtheorem{lemma}[theorem]{Lemma}
\newtheorem{proposition}[theorem]{Proposition}
\theoremstyle{definition}
\newtheorem{definition}[theorem]{Definition}
\theoremstyle{remark}
\newtheorem{remark}[theorem]{Remark}

\begin{document}
%

\title{Widening the light cones on subsets of spacetime: some variations to stable causality}


\author{E. Minguzzi\thanks{
Dipartimento di Matematica Applicata ``G. Sansone'', Universit\`a
degli Studi di Firenze, Via S. Marta 3,  I-50139 Firenze, Italy.
E-mail: ettore.minguzzi@unifi.it} \ and M.
Rinaldelli\thanks{Dipartimento di Matematica ``U. Dini'',  Universit\`a degli Studi di Firenze,
Viale Morgagni 67/A, I-50134 Firenze, Italy. E-mail:
mauro.rinaldelli@math.unifi.it}}

\date{}
\maketitle

\begin{abstract}
\noindent By definition a spacetime is stably causal if it is
possible to widen the light cones all over the spacetime without
spoiling causality. We prove that if the spacetime is at least
non-total imprisoning then it is stably causal provided the light
cones can be widened outside any arbitrarily large compact set, i.e.
in a neighborhood of infinity, without spoiling causality.
Furthermore, we prove that the new causality level `compact stable
causality' can be obtained as the antisymmetry condition of a new
causal relation which we identify, but it cannot be obtained as a
causal stability condition with respect to a  topology on metrics.
The difference between stable causality and compact stable causality
is shown to follow from the fact that Geroch's interval topology on
the space of conformal metrics of $M$
is not Fr\'echet-Urysohn (in fact it is not even
$T$-sequential). In particular we prove that (compact) stably causal
metrics are those in the (sequential) interior of the set of
chronological metrics. Finally, contrary to previous claims it is
shown that stable causality with respect to the $C^0$ fine topology
on metrics leads to the usual notion of stable causality.
\end{abstract}



\section{Introduction}

In the last decades spacetimes have been organized according to
their causality properties in the so-called \emph{causal ladder of
spacetimes}. This ladder is a hierarchy of conformally invariant
properties whose study started at the end of the 60's through the
works of Carter, Geroch, Hawking, Kronheimer, Penrose, Seifert,
Woodhouse and others, who in those years established the main levels
(for an introduction see \cite{hawking73,senovilla97,minguzzi06c}).

One of the most important causality condition is stable causality. A
spacetime is stably causal if the light cones can be widened all
over the spacetime without introducing closed causal curves. For
these spacetimes causality is stable under small perturbations of
the metric. This paper investigates what happens if this condition
is slightly relaxed. This is done in two natural  and complementary
directions. In one case we ask what happens if the enlargment is
done only outside a compact set, namely in a neighborhood of
infinity. In the other we ask what happens if the enlargements have
finite extension, that is, if the widening of the light cones is
done in the interior of a generic compact set.

Concerning the former possibility we prove in section
\ref{Sec_outside} that given a non-total imprisoning spacetime, if
it is possible to widen the light cones outside a compact set
 without spoiling causality, then the spacetime is stably causal
(Theorem \ref{outlight}). In order to prove this result we introduce
a new relation on $M$ whose antisymmetry is tightly connected to our
operation on light cones near infinity. The proof uses the
equivalence between $K$-causality \cite{sorkin96,dowker00} and
stable causality recently proved in \cite{minguzzi08b}.

Concerning the latter possibility we study more in deep {\em compact
stable causality}, a new causality level introduced in
\cite{minguzzi07d} which has been central in order to prove the
mentioned equivalence between $K$-causality and stable causality or
the fact that chronological spacetimes without lightlike lines are
stably causal \cite{minguzzi07d}. A spacetime is compactly stably
causal if it is causal and, roughly speaking, it is possible to open
the light cones over any chosen compact set while preserving
causality. This means that causality is stable under sufficiently
small variations of the metric which are limited in extension.


The definition of compact stable causality is suggested by
variational calculus. Here it is customary to consider metric
variations $\delta g$ with compact support of the Einstein-Hilbert
action in order to get the Einstein equations. If the spacetime is
compactly stably causal, at least for sufficiently small variations,
the corresponding varied spacetimes are all causal. It is
interesting to note that to that end the original spacetime need not
be stably causal as stable causality differs from compact stable
causality \cite{minguzzi07d}.
%

Many causality conditions are traced back to antisymmetry conditions
 on causality relations so that the relationship between the different
causality requirements becomes trivial  \cite{minguzzi07b} and
related to the inclusion of sets on $M \times M$. For instance,
Seifert introduced \cite{seifert71} the relation
$J_S^+=\cap_{g'>g}{J}^+_{g'}$ and proved its transitivity and
closure. He also argued that $J_S^+$ is antisymmetric if and only if
the spacetime is stably causal (for a rigorous proof see \cite[Proposition
2.3]{hawking74} or \cite[Theorem 3.12]{minguzzi07}). It must be
recalled here that the usual causal relation $J^{+}$ although
transitive is not closed, a fact which explains why $J^{+}_S$ is
particularly interesting. About twelve years ago, Sorkin and Woolgar
reconsidered the properties of closure and transitivity but took a
different approach. They defined the relation $K^+ \subset M \times
M$ as the smallest closed and transitive relation containing $J^+$,
moreover, they defined a spacetime to be $K$-causal if the relation
$K^+$ is antisymmetric \cite{sorkin96}.

Since compact stable causality is  similar to stable causality we
expect to find a causal relation that plays  for compact stable
causality the same role that $J^{+}_S$ plays for stable  causality.
In { section \ref{Sec_CSC}} we identify such new causality relation
and prove that its antisymmetry is  necessary  and sufficient  for
the compact stable causality of the spacetime (Theorem \ref{Jcs}).

In Table \ref{tabella} we summarize the portion of the causal ladder
below stable causality. In this figure whenever possible we provide
the causal relation whose antisymmetry determines the level along
with its closure and transitivity properties. For a unified
framework showing all the causal relations that have appeared so far
in the literature see \cite{minguzzi07b,minguzzi07}.

\begin{table}
    \centering
        \begin{tabular}{|c|c|c|c|}
          \hline
            {\small  Causal Ladder} & {\small Antisymmetry of relation } & {\small Transitive} & {\small Closed} \\
            \hline
            & & & \\
            {\small Stable causality} &  {\small $J^+_S=\bigcap_{g'>g} \! J^+_{g'}$} & {\small yes} & {\small yes} \\
            {\small $\Updownarrow$} & & & \\
            {\small $K$-causality} & {\small $K^+$} & {\small yes } & {\small yes} \\
            {\small $\Downarrow$} & & &\\
            {\small $\overline{A^{\infty}}$-causality} & {\small $\overline{A^{+\infty}}$} & {\small no}  & {\small yes}\\
            {\small $\Downarrow$} & & & \\
            {\small Compact stable causality} & {\small $J^+_{CS} = \bigcup_{B} \bigcap_{g'\in \{g\}_B} \! J^+_{g'}$ } & {\small yes } & {\small no}\\
            {\small $\Downarrow$} & & & \\
            {\small $A^{\infty}$-causality} & {\small $A^{+\infty}=\cup^{+\infty}_{i=1}(A^+)^i$ } & {\small yes}  &  {\small no}\\
            {\small $\Downarrow$} & & &\\
            {\small $A$-causality} & {\small $A^+=\bar{J}^+$} &{\small no} & {\small yes} \\
            {\small $\Downarrow$} & & &\\
            {\small Strong causality} & -- & -- & -- \\
            {\small $\Downarrow$} & & & \\
            {\small Non-partial imprisonment} & -- & --  & -- \\
            {\small $\Downarrow$} & & & \\
             {\small Weak distinction } & {\small $D^{+}$} & {\small yes} & {\small no} \\
            {\small $\Downarrow$} & & &\\
            {\small Non-total imprisonment} & -- & -- & -- \\
            {\small $\Downarrow$}& & & \\
            {\small Causality} & {\small $J^+$} & {\small yes} & {\small no} \\
            {\small$\Downarrow$} & & & \\
            {\small Chronology} & {\small $I^+$} & {\small yes} & {\small no} \\
            \hline
        \end{tabular}
    \label{tabella}
    \caption{The causal ladder and, for each level, the corresponding causal relation whose
    antisymmetry determines the causality condition. For the definition of $D^{+}$ see \cite{minguzzi07e}.
    The last two columns report on the  transitivity and closure of the relation in the most general case; they
    can be both {\em yes} for particular spacetimes.}
\end{table}

 In section \ref{Sec_topologies} we study stable
causality and compact stable causality in their relationship with
the possible topologies on the space of Lorentzian metrics. We first
show which natural topologies lead to stable causality, then we
follow a reasoning which argues that compact stable causality can
not be obtained as a causal stability condition with respect to a
reasonable topology on metrics. Finally, we focus on Geroch's
interval topology and show that there is a neat topological
connection between stable causality and compact stable causality which
resides in the difference between the interior and the sequential
interior of a set. Indeed in the interval topology these two
interior concepts differ as the topology is not Fr\'echet-Urysohn.
In theorems \ref{Int} and \ref{Ints} we prove that the stably causal
metrics are those which stay in the interior of the set of
chronological metrics, while the compactly stably causal metrics are
those in the sequential interior.

We refer the reader to \cite{minguzzi07b,minguzzi06c} for most of
the conventions used in this work. In particular, we denote with
$(M,g)$ a $C^r$ spacetime (connected, time-oriented Lorentzian
manifold), $r \in {3, \dots, \infty}$ of arbitrary dimension $n \geq
2$ and signature $(-,+, \dots, +)$. On $M \times M$ the usual
product topology is defined. The subset symbol $\subset$ is
reflexive, thus $X \subset X$. With $J^+_{g}$ we specify the causal
relation referring to metric $g$.

$\textrm{Lor}(M)$ denotes the space of all Lorentzian metrics for a
given manifold $M$; a partial ordering may be defined on
$\textrm{Lor}(M)$ by $g_1<g_2$ if $g_1(v,v)\leq 0$ implies
$g_2(v,v)<0$ for all $v\neq 0$ in $TM$. $\textrm{Con}(M)$ denotes
the quotient space formed by identifying all pointwise globally
conformal metrics $g_1=\Omega g_2$, with $\Omega:M \rightarrow
(0,\infty)$ smooth. With $[g]$ it is denoted the conformal class of
$g$. Let $g_1,g_2 \in \textrm{Lor}(M)$, $[g_1],[g_2] \in
\textrm{Con}(M)$. Let $g_1' \in [g_1]$, $g_2' \in [g_2]$, be
alternative representative, since $g_1<g_2$ iff $g_1'<g_2'$ then the
partial ordering on $\textrm{Lor}(M)$ may be projected naturally to
a partial ordering on $\textrm{Con}(M)$. We shall therefore write
$[g_1]<[g_2]$ or simply $g_1<g_2$ being clear from the context if
with the symbol $g$ it is understood a metric or a conformal class.
In this article we will mostly handle conformal invariant
properties, thus $(M,g)$ is usually used with the meaning of
$(M,[g])$, or better $[(M,g)]$, the class of spacetimes with
conformal metrics and the same time-orientation (for a rigorous
definition see \cite{minguzzi06c}).

\section{Widening the light cones outside a compact set} \label{Sec_outside}

We define $g_1\prec g_2$ if $g_1 \le g_2$ and $g_1 <g_2$ except over a
compact (possibly empty) set where the equality may hold. If $C$ is
a compact set we shall also write $g_1\prec_{{}_C} g_2$ if $g_1=
g_2$ on $C$ and $g_1<g_2$ outside $C$.

We want to prove that if a spacetime is non-totally imprisoning and
non-stably causal, then for every $\tilde{g} \succ g$,
$(M,\tilde{g})$ is not-causal, i.e. there exists a closed
$\tilde{g}$-causal curve. As we shall see the proof uses the
equivalence between $K$-causality and stable causality as recently
proved in \cite{minguzzi08b}.


As a first step we introduce the new relation
\[
R^+ =  \bigcap_{\tilde{g} \succ g} J^+_{\tilde{g}}.
\]

Note that we can also write $ R^{+}=\bigcap_C \bigcap_{\tilde{g}
\succ_{\!\!{}_C} g} J^+_{\tilde{g}} $
 where the first intersection is over
the set of compact sets.

The idea is to prove that $R^+$ is antisymmetric if and only if
there is $\tilde{g} \succ g$ such that $(M,\tilde{g})$ is causal, a
fact which will be used in the proof of the main thesis above. We
will prove this statement later.

First, note that $J^{+}\subset R^+$ because $J^{+}\subset
J^+_{\tilde{g}}$ for every $\tilde{g} \succ g$. In particular
$R^{+}$ is non-empty. Let us investigate the closure and
transitivity properties of $R^+ \subset M \times M$.

\begin{lemma} \label{trans}
$R^+$ is transitive.
\end{lemma}
\begin{proof}
If $(x,y) \in R^+$ and $(y,z) \in R^+$, then for every compact set
$C$ and for every $g' \succ_{\!\!{}_C} g$, $(x,y) \in J^+_{g'}$ and
$(y,z) \in J^+_{g'}$. Since $J^+_{g'}$ is transitive, $(x,z) \in
J^+_{g'}$ and since $C$ and $g' \succ_{\!\!{}_C} g$ are arbitrary,
$(x,z) \in R^+$  and thus $R^{+}$ is transitive.
\end{proof}

\begin{lemma} \label{clos}
If the spacetime is non-totally imprisoning, then
$$
\bigcap_C \bigcap_{\tilde{g} \succ_{\!\!{}_C} g}
\bar{J}^+_{\tilde{g}} = \bigcap_C \bigcap_{\tilde{g}
\succ_{\!\!{}_C} g}  J^+_{\tilde{g}} .
$$

\end{lemma}
\begin{proof}
In one direction the inclusion is trivial. Let  $C \subset M$ be an
arbitrary compact set and consider an arbitrary metric   $\hat{g}
\succ_{\!\!{}_C} g$.  Consider an arbitrary pair $(x,y) \in
\bigcap_K \bigcap_{\tilde{g} \succ_{\!\!{}_K} g
}\bar{J}^+_{\tilde{g}}$, we are going to prove that $(x,y) \in
J^+_{\hat{g}}$. In fact, let $g'$ such that
$\hat{g}\succ_{\!\!{}_C}g'\succ_{\!\!{}_C}g$, then there are two
cases:
\begin{itemize}
    \item[(i)] $(x,y) \in J^+_{g'}$ thus $(x,y) \in J^+_{\hat{g}}$;
    \item[(ii)] $ (x,y) \notin J^+_{g'}$ but since we know that $ (x,y) \in \bigcap_K \bigcap_{\tilde{g}
    \succ_{\!\!{}_K} g}
\bar{J}^+_{\tilde{g}}\subset \bar{J}^+_{g'}$,  by the limit curve
theorem \cite[Theorem 3.1]{minguzzi07c} there are a future
inextendible $g'$-causal curve $\sigma^x$ starting from $x$ and a
past inextendible $g'$-causal curve $\sigma^y$ ending at $y$. Since
the spacetime is non-totally imprisoning,  both $\sigma^x$ and
$\sigma^y$ escape $C$. Let $x' \in (\sigma^x\backslash\{x\}) \cap
(M\backslash C)$ and $y' \in (\sigma^y\backslash\{y\}) \cap
(M\backslash C)$, by the limit curve theorem $(x',y') \in
\bar{J}^+_{g'}$. Since the segment of $\sigma^x$ between $x$ and
$x'$ intersects the open set $M\backslash C$, where $g'<\hat{g}$, we
have  $(x,x') \in I^+_{\hat{g}}$ and analogously $(y',y) \in
I^+_{\hat{g}}$. Since $I^+_{\hat{g}}$ is open, and $\bar{J}^+_{g'}
\subset \bar{J}^+_{\hat{g}}$ these relations imply $(x,y) \in
J^+_{\hat{g}}$.
\end{itemize}

Since $C$ and $\hat{g}$ are arbitrary, $(x,y) \in \bigcap_C
\bigcap_{\hat{g} \succ_{\!\!{}_C} g }J^+_{\hat{g}}$ from which the
thesis follows.
\end{proof}

\begin{corollary}
If the spacetime is non-totally imprisoning, then $R^{+}$ is closed.
\end{corollary}

\begin{proof}
Since $R^{+}$ is the intersection of closed sets, $R^+ =  \bigcap_C
\bigcap_{\tilde{g}
    \succ_{\!\!{}_C} g} \bar{J}^+_{\tilde{g}}$, it is closed.
\end{proof}

Recall that $J^+_S=\bigcap_{g'>g} J^+_{g'}$ is the Seifert relation
\cite{seifert71}, and that $K^+$ is the smallest relation on $M$
which contains $I^+$ and is closed and transitive. The relation
$J^+_S$ is closed and transitive
\cite{seifert71,hawking74,minguzzi07}, thus $K^+ \subset J^+_S$, and
in \cite{minguzzi08b} it has been proved that if a spacetime is
$K$-causal, then $K^+ = J^+_S$.

\begin{lemma} \label{lemma3}
If the spacetime is non-totally imprisoning, then $K^+ \subset R^+
\subset J^+_S$.
\end{lemma}
\begin{proof}
The inclusion $K^+ \subset R^+$ follows immediately from the fact
that $R^+$ is transitive and closed under non-total imprisonment,
and that $K^+$ is the smallest set with these properties.

The inclusion $R^+ \subset J^+_S$ is trivially true because the set
of metrics over which we take the intersection in the definition of
$R^{+}$ is larger than that for $J^+_S$ (in the definition of
$J^{+}_S$ the metrics in the intersection coincide with $g$ in a
compact set, namely the empty set).
\end{proof}

%
%
%
%

\begin{lemma} \label{lemma4}
Let $(M,g)$ be a non-totally imprisoning spacetime. The following properties are equivalent:
\begin{itemize}
\item[(i)] The relation
$R^+$ is antisymmetric.
\item[(ii)] There is $\tilde{g}
    \succ g$  such that $(M,\tilde{g})$ is causal.
\item[(iii)] The spacetime is stably causal.
\end{itemize}
\end{lemma}

\begin{proof}

(i) $\Rightarrow$ (iii). Since $R^{+}$ is antisymmetric, by lemma \ref{lemma3}
$\; K^{+}$ is antisymmetric, thus by the result of \cite{minguzzi08b}
$\; K^{+}=R^{+}=J^{+}_S$, and in particular the spacetime is stably
causal. \\
(iii) $\Rightarrow$ (ii). There is $\tilde{g}>g$ such that $(M,\tilde{g})$ is
causal and note that $\tilde{g} \succ g$ as they coincide only over
a compact set (the empty set). \\
(ii) $\Rightarrow$ (i). (non-total imprisonment is not used) Let $x,y \in M$,
 such that $(x,y) \in R^+, (y,x) \in R^+$, thus for the
metric $\tilde{g}
    \succ g$ of the hypothesis,
 $(x,y) \in J^+_{\tilde{g}}$ and
$(y,x) \in J^+_{\tilde{g}}$. As  $(M,\tilde{g})$ is causal $x=y$
thus $R^{+}$ is antisymmetric.

\end{proof}

\begin{theorem} \label{outlight}
If $(M,g)$ is non-totally imprisoning  but non-stably causal, then
for every $\tilde{g}\succ g$, there exists a closed
$\tilde{g}$-causal curve.

More strongly, if $(M,g)$ is non-totally imprisoning  but non-stably
causal, then there are $x,y \in M$, $x \ne y$ such that for every
$\tilde{g}\succ g$, there exists a closed $\tilde{g}$-causal curve
passing through $x$ and  $y$.
\end{theorem}
\begin{proof}
Since $(M,g)$ is not stably causal $R^{+}$ is not antisymmetric,
thus there are $x,y \in M$, $x\ne y$, $(x,y)\in R^{+}$ and $(y,x)\in
R^{+}$, thus for every $\tilde{g} \succ g$, $(x,y)\in
J^{+}_{\tilde{g}}$ and $(y,x)\in J^{+}_{\tilde{g}}$.

\end{proof}

\begin{figure}[htbp]
    \centering
        \includegraphics[scale=.9]{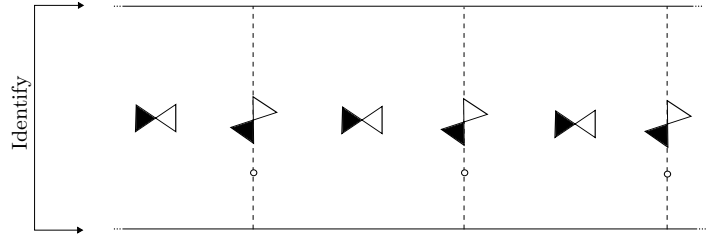}
    \caption{The picture continues indefinitely in the horizontal direction, it displays a non-total imprisoning spacetime such that no matter the chosen compact set $C$, $(M\backslash C,g)$ is non-stably
causal. }
    \label{fig:VerticalStrip}
\end{figure}

The previous result shows that a widening of the light cones near
infinity produces closed causal curves which pass always through some
 points no matter how much this widening is made `close to
infinity'. The pathological behavior has to be attributed to the
spacetime `at infinity': indeed, by removing an arbitrarily large
compact set one cannot cure this problem. The example of
figure \ref{fig:VerticalStrip} gives a non-total imprisoning
spacetime such that, no matter the compact  set $C$, $(M\backslash
C,g)$ is non-stably causal.

\section{Compact stability relation and compact stable causality} \label{Sec_CSC}

In this section we make large use of metrics which are widened in a
specified set, so we find useful to introduce a new notation: if $B$
is a  relatively compact open set and $g$ the original metric of the
spacetime, we denote with $\{g\}_B$ the family of metrics $g'\geq g$
such that $g'>g$ on $B$ and $g'= g$ on $M \backslash B$. Moreover,
if $C$ is a \emph{compact} set, then $\{g\}_C$ is the family of
metrics $g'\geq g$ such that $g'>g$ on $C$.

We start by giving alternative definitions of compact stable
causality \cite{minguzzi07d}

\begin{definition} \label{CSC}
A spacetime $(M,g)$ is \emph{compactly stably causal} if one of the following equivalent properties hold:
\begin{itemize}
    \item[(i)] for every relatively compact open set $B$ there is a metric $\tilde{g}_B\geq g$ such that $\tilde{g}_B>g$ on $B$, $\tilde{g}_B=g$ on $M\backslash B$ and $(M,\tilde{g}_B)$ is causal.
    \item[(ii)] for every relatively compact open set $B$ there is a metric $g_B\geq g$ such that $g_B>g$ on $B$ and $(M,g_B)$ is causal.
    \item[(iii)] for every compact set $C$ there is $g_C\geq g$ such that $g_C>g$ on $C$ and $(M,g_C)$ is causal.
\end{itemize}
\end{definition}

\begin{proof}[Proof of the equivalence]
(i) $\Rightarrow$ (ii). Take $g_B=\tilde{g}_B$. (ii)$\Rightarrow$
(i). Take a convex combination of $g_B$ with with $g$, $\tilde{g}_B
= \chi g_B + (1-\chi)g$ where $\chi: M \to [0,1]$ is a function
which is positive in $B$ and vanishes outside $B$. Since
$\tilde{g}_B \leq g_B$, $(M,\tilde{g}_B)$ is causal. (ii)
$\Rightarrow$ (iii). Take $B$ such that $C \subset B$ and define
$g_C=g_B$. (iii) $\Rightarrow$ (ii). Take $C=\bar{B} $.
\end{proof}

Compact stable causality finds place in the causal ladder between
$A^\infty$-causality and $\overline{A^\infty}$-causality, as proved
in \cite{minguzzi07d}.

Recall that \cite{minguzzi07},  $g\leq g' \Rightarrow J^+_g\subset
J^+_{g'}$, and $g<g' \Rightarrow \bar{J}^+_g\subset J^+_{g'}$.

\begin{definition}
$J^+_{CS}$ is the subset of $M \times M$ defined by

\begin{equation} J^+_{CS} = \bigcup_{B}  \!  \bigcap_{g' \in \,\{g\}_B} \! \! \! \!
J^+_{g'},
\end{equation}
 where $B$ ranges over all relatively compact
open sets. The index \emph{CS} stands for \emph{Compact Stability
relation}.
\end{definition}

\begin{proposition} We have
\[\bigcup_{C}  \!  \bigcap_{g' \in\, \{g\}_C} \! \! \!  J^+_{g'} \; = \;
\bigcup_{B}  \!  \bigcap_{g' \in \,\{g\}_B} \! \! \!  J^+_{g'}, \]
where $C$ ranges over all compact sets and $B$ ranges over all
relatively compact open sets. Thus it is possible to give the
alternative definition $J^+_{CS} =\bigcup_{C}  \!  \bigcap_{g' \in\,
\{g\}_C} \! \! \!  J^+_{g'}$.
\end{proposition}

\begin{proof}

Let $C$ be a compact set and let $B$ be a relatively compact open
set such that $B\supset C$. We show that
\[
\bigcap_{g_C \in \{g\}_C} \!\!\!\! J^+_{g_C}\subset J^+_{g_B} \quad
\quad \forall g_B \in \{g\}_B\,.
\]
 Indeed, whatever $g_B \in
\{g\}_B$ we can find $\bar{g}_C \in \{g\}_C$ such that $g \leq
\bar{g}_C \leq g_B$, and then $J^+_{\bar{g}_C} \subset J^+_{g_B}$ so
that $\bigcap_{g_C \in \{g\}_C} J^+_{g_C} \subset J^+_{g_B}$. Since
$g_B \in \{g\}_B$ is arbitrary $\bigcap_{g_C \in \{g\}_C} J^+_{g_C}
\subset \bigcap_{g_B \in \{g\}_B} J^+_{g_B}$, thus $\bigcap_{g_C \in
\{g\}_C} J^+_{g_C} \subset\bigcup_{B} \bigcap_{g_B \in \{g\}_B}
J^+_{g_B}$ and finally  $\bigcup_{C} \bigcap_{g_C \in \{g\}_C}
J^+_{g_C} \subset \bigcup_{B}\bigcap_{g_B \in \{g\}_B} J^+_{g_B}$.

For the converse let $B$ be a relatively compact open set and let
$C$ be a compact set such that  $C \supset \bar{B}$. We show that
\[
\bigcap_{g_B \in \{g\}_B} \!\!\!\! J^+_{g_B}\subset J^+_{g_C} \quad
\quad \forall g_C \in \{g\}_C\, .
\]
Indeed, whatever is $g_C \in \{g\}_C$ we can find $\bar{g}_B \in
\{g\}_B$ such that $g \leq \bar{g}_B \leq g_C$, and thus
$J^+_{\bar{g}_B} \subset J^+_{g_C}$ so that $\bigcap_{g_B \in
\{g\}_B} J^+_{g_B} \subset J^+_{g_C}$. Since $g_C \in \{g\}_C$ is
arbitrary $\bigcap_{g_B \in \{g\}_B} J^+_{g_B} \subset \bigcap_{g_C
\in \{g\}_C} J^+_{g_C}$, thus $\bigcap_{g_B \in \{g\}_B} J^+_{g_B}
\subset \bigcup_C \bigcap_{g_C \in \{g\}_C} J^+_{g_C}$, and finally
 \[\bigcup_{B} \bigcap_{g_B \in \{g\}_B} J^+_{g_B} \subset
\bigcup_{C} \bigcap_{g_C \in \{g\}_C} J^+_{g_C}.\]

\end{proof}

\begin{proposition} \label{transitivity}
$J^+_{CS}$ is transitive.
\end{proposition}
\begin{proof}
$(x,y)\in J^+_{CS}$ means that there is a compact set $C_{xy}$ such
that for every $g'\ge g$,  $g'>g$ in $C_{xy}$, it is $(x,y) \in
J^{+}_{g'}$. Analogously, $(y,z)\in J^+_{SC}$ means that there is a
compact set $C_{yz}$ such that for every $g'\ge g$, $g'>g$ in
$C_{yz}$, it is $(x,y) \in J^{+}_{g'}$.  Consider
$C_{xz}:=C_{xy}\cup C_{yz}$ then for every $g'\ge g$, $g'>g$ in
$C_{xz}$, it is in particular $g'>g$ both in $C_{xy}$ and in
$C_{yz}$, thus due to the transitivity of $J^{+}_{g'}$, $(x,z) \in
J^{+}_{g'}$.
\end{proof}

The remainder of the section is devoted to the proof of the
equivalence between compact stable causality and the antisymmetry of
$J^+_{CS}$. In one direction the proof is simple

\begin{lemma} \label{oneside}
If $(M,g)$ is compactly stably causal then $J^+_{CS}$ is
antisymmetric.
\end{lemma}
\begin{proof}
Assume that $J^+_{CS}$ is not antisymmetric, then there are $x,z$,
$x\ne z$, such that $(x,z) \in J^+_{CS}$ and $(z,x) \in J^+_{CS}$.
Thus there is a compact set $C_{xz}$ such that for every metric
$g_{C_{xz}} \in \{g\}_{C_{xz}}$, $(x,z) \in J^+_{g_{C_{xz}}}$,
analogously there is a compact set $C_{zx}$ such that for every
metric $g_{C_{zx}} \in \{g\}_{C_{zx}}$, $(z,x) \in
J^+_{g_{C_{zx}}}$. As a consequence for the compact set $C=C_{xz}
\cup C_{zx}$, every metric $g_C \in \{g\}_C$ can be considered as a
metric belonging to $\{g\}_{C_{xz}}$ and $\{g\}_{C_{zx}}$ thus
$(x,z) \in J^+_{g_C}$ and $(z,x) \in J^+_{g_C}$ thus $(M,g)$ is not
compactly stably causal.
\end{proof}

The proof of the converse, that is that the antisymmetry of
$J^+_{CS}$ implies compact stable causality, is more complex.
Indeed, we shall need some preliminary lemmas. The overall strategy
will be close to that of \cite{minguzzi07} where it is proven that
stable causality is equivalent to the antisymmetry of the Seifert's
relation $J_S^+$.

We know from lemma 3.2 of \cite{minguzzi07} that \emph{if
$\tilde{g}<g$ then $\bar{J}^+_{\tilde{g}} \subset \Delta \cup I^+_g
\subset J^+_g$}. An analogous result holds if the two metrics
coincide outside a compact set

\begin{lemma} \label{3.2}
Let $B$ be a relatively compact open set. If $\tilde{g}<g$ in $B$
and $\tilde{g}=g$ in $M \backslash B$, then $\bar{J}^+_{\tilde{g}}
\cap (B\times B)\; \subset \; (\Delta \cup I^+_g) \cap (B\times
B)\subset \; J^+_g \cap (B\times B)$.
\end{lemma}

\begin{proof}
Let $(x,z) \in (\bar{J}^+_{\tilde{g}} \backslash \Delta) \cap
(B\times B)$, let $\sigma_n$ be a sequence of ($\tilde{g}$-)causal
curves of endpoints $(x_n,z_n) \to (x,z)$. If $(x,z)\in
J^+_{\tilde{g}}$ then the ($\tilde{g}$-)causal curve which connects
$x$ to $z$ necessarily intersects $B$, thus $(x,z)\in I^{+}_{g}$. We
can therefore assume $(x,z)\notin J^+_{\tilde{g}}$. Using the limit
curve theorem \cite{minguzzi07c} it follows the existence of a
future inextendible ($\tilde{g}$-)causal curve $\sigma^x$ starting
from $x$, a past inextendible ($\tilde{g}$-)causal curve $\sigma^z$
ending at $z$, and a subsequence $\sigma_j$ distinguishing both
curves. Taken $x' \in (\sigma^x\backslash\{x\}) \cap B$, $z' \in
(\sigma^z\backslash\{z\}) \cap B$  it follows $(x,x') \in
J^+_{\tilde{g}} \cap (B \times B)$, $(z',z) \in J^+_{\tilde{g}} \cap
(B \times B)$ and $(x',z') \in \bar{J}^+_{\tilde{g}} \cap (B \times
B)$. In terms of the causal relations of $(M,g)$, since the piece of
$\sigma^x$ between $x$ and $x'$ intersects $B$, it is $(x,x') \in
I^+_g$, and analogously $(z',z) \in I^+_g$. Moreover, $(x',z') \in
\bar{J}^+_g$, which implies, because $I^+_g$ is open, $(x,z) \in
I^+_g$.
\end{proof}

\begin{lemma} \label{3.3}
Let $B$ be a relatively compact open set, then
$$\bigcap_{g_B \in \{g\}_B} \!\!\!\! J^+_{g_B} \cap (B \times B) \; =  \bigcap_{g_B \in \{g\}_B} \!\!\!\! \bar{J}^+_{g_B} \cap (B \times B).$$
\end{lemma}
\begin{proof}
We have only to show that
\[
\bigcap_{g_B \in \{g\}_B} \!\!\!\! \bar{J}^+_{g_B} \cap (B \times B)
\quad \subset  \bigcap_{g_B \in \{g\}_B} \!\!\!\! J^+_{g_B} \cap (B
\times B),
\] the other inclusion
being obvious. Let $\bar{g} \in \{g\}_B$, taken $\tilde{g} \in
\{g\}_B$ such that $g<\tilde{g}<\bar{g}$ in $B$, by  lemma \ref{3.2}
it is $\bar{J}^+_{\tilde{g}} \cap (B\times B)\; \subset \;
J^+_{\bar{g}} \cap (B\times B)$, thus $\bigcap_{g_B \in \{g\}_B}
\bar{J}^+_{g_B} \cap (B \times B) \; \subset J^+_{\bar{g}} \cap (B
\times B)$. Since $\bar{g}$ is arbitrary the thesis follows.
\end{proof}

Recall that a spacetime is chronological at $x$ if no closed
timelike curve passes through $x$.

\begin{lemma} \label{3.9}
If $J^+_{CS}$ on $(M,g)$ is antisymmetric then for every relatively
compact open set $V \subset M$ and for every $x \in V$
there is a ($x$-dependent) metric $g_x \in \{g\}_{V}$ such that $(M,g_x)$ is
chronological at $x$.
\end{lemma}
\begin{proof}
Assume by contradiction that the thesis does not hold, then there is
a relatively compact open set $V$ and some $x \in V$ such that for
every $g' \in \{g\}_{V}$ there is a closed ($g'$-)timelike curve
passing through $x$. Fix a $\bar{g} \in \{g\}_{V}$, introduce a
Riemannian metric in a neighborhood of $x$ and consider
$S=\dot{B}(x,\epsilon)$, i.e. the surface of the ball of Riemannian
radius $\epsilon >0$. Choose $\epsilon$ sufficiently small so that
$S$ is contained in a ($\bar{g}$-)convex neighborhood contained in a
($\bar{g}$-)globally hyperbolic neighborhood $W$ contained in $V$.

For every $g' \in \{g\}_{V}$, $g<g'<\bar{g}$ in $V$, there is a
closed ($g'$-)timelike curve $\sigma_{g'}$ passing through $x$. This
curve must escape the hyperbolic neighborhood $W$ otherwise in
$(W,\bar{g})$ there would be a closed ($\bar{g}$-)timelike curve.
Hence the curve must meet $S$ at some point of $S \cap
I^+_{\bar{g}}(x)$. Given $g'$ the event $x$ belongs to the
\emph{chronology violating set $vI_{g'}$} which is open
\cite{penrose72} and which can be written as the union of disjoint
open sets of the form $I^+_{g'}(y) \cap I^-_{g'}(y)$ where $y$ is
any point of the component \cite[Proposition 4.27]{penrose72}. In
particular $x$ belongs to the component $I^+_{g'}(x) \cap
I^-_{g'}(x)$. The set $A(g')=I^+_{g'}(x) \cap I^-_{g'}(x) \cap S
\cap I^+_{\bar{g}}(x)\neq \emptyset$ is open in the topology
inherited by $S$ and non empty because $\sigma_{g'}$ must meet
$S\cap I^+_{\bar{g}}(x)$. In the topology of $S$, $\bar{A}(g')$ are
non-empty compact sets, thus $\bigcap_{g'} \bar{A}(g') \neq
\emptyset$, where the intersection is taken over all $g' \in
\{g\}_{V}$ such that $g<g'<\bar{g}$ in $V$ (this result follows from
Cantor's intersection lemma \cite[Theorem 3.1.1]{engelking89}, and
the fact that the family $\{\bar{A}(g')\}$ has the finite
intersection property, for more details see the proof of Lemma 3.9
in \cite{minguzzi07}). As a consequence $\bigcap_{g'\in \{g\}_{V}}
\bar{A}(g') \neq \emptyset$ and hence there is $z \in \bigcap_{g'\in
\{g\}_{V}} \bar{A}(g') \neq \emptyset$.

In other words there is an event $z \in S$ such that for every $g'
\in \{g\}_{V}$, $g<g'<\bar{g}$ in $V$, there are closed
($g'$-)timelike curves starting from $x$ and passing arbitrarily
close to $z$. Thus for every $g' \in \{g\}_{V}$, $(x,z) \in
\bar{J}^+_{g'}$ and $(z,x) \in \bar{J}^+_{g'}$, thus by lemma
\ref{3.3} $(x,z) \in \bigcap_{g' \in \{g\}_{V}} J^+_{g'}$ and $(z,x)
\in \bigcap_{g' \in \{g\}_{V}} J^+_{g'}$, so $(x,z) \in J^+_{CS}$
and $(z,x) \in J^+_{CS}$; but $x \neq z$, i.e. $J^+_{CS}$ is not
antisymmetric.
\end{proof}

Recall that a spacetime is strongly causal  at $x$ if it admits
arbitrarily small causally convex neighborhoods of  $x$.

\begin{lemma} \label{3.10_bis}
Let $B$ be a relatively compact open set. If $(M,g)$ is
chronological at $x \in B$ then for every $g'$ such that $g'<g $ in
$B$ and $g'=g$ in $M \backslash B$, $(M,g')$ is strongly causal at
$x$. (Stated in another way, if $(M,g')$ is non-strongly causal at
$x \in B$ then for every $g$ such that $g>g' $ in $B$ and $g=g'$ in
$M \backslash B$ there is a ($g$-)timelike closed curve passing
through $x$.)
\end{lemma}

\begin{proof}
If $(M,g')$ is not strongly causal at $x$ then the characterizing
property (ii) of \cite[Lemma 3.21]{minguzzi06c} does not hold, that
is, there is a neighborhood $U\ni x$ and a sequence of ($g'$-)causal
curves $\sigma_n$ of endpoints $x_n,z_n$, with $x_n \rightarrow x$,
$z_n \rightarrow x$, not entirely contained in $U$. Let $C \ni x$ be
a ($g'$-)convex neighborhood whose compact closure is contained in
another ($g'$-)convex neighborhood $V \subset (U \cap B)$  (they
exist, see \cite{penrose72} or \cite{minguzzi06c}). Let $c_n \in
\dot{C}$ be the first point at which $\sigma_n$ escapes $C$, and let
$d_n$ be the last point at which $\sigma_n$ reenters $C$. Since
$\dot{C}$ is compact there are $c,d \in \dot{C}$, and a subsequence
$\sigma_k$ such that $c_k \rightarrow c$, $d_k \rightarrow d$ and
since $V$ is convex, the causal relation on $V \times V,
J^+_{(V,g')}$, is closed and hence $(x,c), (d,x) \in J^+_{(V,g')}$
thus $(x,c), (d,x) \in J^+_{g'}$ (note that $d$ and $c$ must be
distinct since the spacetime $(V,g')$ is causal as $V$ is convex).
Taking into account that $(c_k,d_k) \in J^+_{g'}$ it is $(c,d) \in
\bar{J}^+_{g'}$. Thus, switching to $g\geq g'$ as in the statement
of this lemma, there is a ($g$-)timelike curve connecting $d$ to $c$
passing through $x$, and since $I^+_g$ is open this is also true for
two neighborhoods of $d$ and $c$. Now, being $(c,d) \in \bar{J}^+_g$
there is a closed ($g$-)timelike curve passing through $x$.
\end{proof}

In other words this lemma  states that if we have chronology at an
event $x$, we can obtain strong causality by narrowing the light
cones in any chosen neighborhood of $x$.

\begin{lemma} \label{3.11_bis}
If for every relatively compact open set $V \subset M$ and for every
$x \in V$ there is a ($x$ dependent) $g_x \in \{g\}_V$ such that
$(M,g_x)$ is chronological at $x$ then $(M,g)$ is compactly stably
causal. (Stated in another way, if $(M,g)$ is non-compactly stably
causal then there exist $V$ relatively compact open set and an event
$x \in V$ such that for every $\bar{g}\in \{g\}_V$, $(M,\bar{g})$ is
non-chronological at $x$).
\end{lemma}

\begin{proof}
Using the second statement, let $(M,g)$ be non-compactly stably
causal, i.e. there exists $B$ relatively compact open set such that
for every $\bar{g} \in \{g\}_B$ there is a closed $\bar{g}$-causal
curve. Assuming $(M,g)$ causal (otherwise the theorem is trivially
true), every such closed $\bar{g}$-causal  curve passes through $B$.

Let $V \supset \bar{B}$ be a relatively compact open set, then for
every $\tilde{g} \in \{g\}_V$ there exists a $\tilde{g}$-causal
closed curve: indeed, for every $\tilde{g} \in \{g\}_V$ there exists
a $\bar{g} \in \{g\}_B$ such that $ \bar{g} \leq  \tilde{g}$.

Now, if the thesis weren't true, for every $y \in V$ there would be
$\tilde{g}_y \in \{g\}_V$ such that $(M,\tilde{g}_y)$ is
chronological in $y$. By lemma \ref{3.10_bis}, taken $g_y$ such that
$g<g_y<\tilde{g}_y$ on $V$, $(M,g_y)$ is strongly causal at $y$ and
hence it is strongly causal in an open neighborhood $U_y$ of $y$
\cite{penrose72}.

From the open covering $\{U_y,y \in \bar{B}\}$, for the compact set
$\bar{B}$ a finite covering can be extracted $\{U_{y_1}, U_{y_2},
\dots, U_{y_k}\}$, and a metric $g^* \in \{g\}_V$ can be found such
that for $i=1,\dots,k$, $g^*<g_{y_i}$ on $V$. Thus, $(M,g^*)$ is
still strongly causal on an open set $A=\bigcup_i U_{y_i} \supset
\bar{B}$. Let $\chi_B:M\rightarrow [0,1]$ be a smooth function such
that $\chi_B=0$ outside $B$ and $g'=(1-\chi_B)g+\chi_Bg^*$. It is
$g' \in \{g\}_B$ by construction; furthermore $g'\leq g^*$ and hence
$(M,g')$ is  causal at every point of $B$ and hence on $M$, a
contradiction with the hypothesis.
\end{proof}

\begin{theorem} \label{Jcs}
The relation $J^+_{CS}$ on $M \times M$ is antisymmetric if and
only if $(M,g)$ is compactly stably causal.
\end{theorem}
\begin{proof}
We have already proved (lemma \ref{oneside})  that compact stable
causality implies the antisymmetry of $J^+_{CS}$.

For the converse let $J^+_{CS}$ be antisymmetric, then for every
relatively compact open set $V \subset M$ and for every
$x \in V$ there is (lemma \ref{3.9}) a $x$-dependent metric $g_x \in \{g\}_{V}$ such
that $(M,g_x)$ is chronological at $x$, thus $(M,g)$ is compactly
stably causal because of lemma \ref{3.11_bis}.
\end{proof}


\section{Topologies on the space of Lorentzian metrics} \label{Sec_topologies}

In \cite{hawking71} Hawking introduces three kind of $C^0$
topologies on the space $\textrm{Lor}(M)$ of the Lorentzian metrics
$g$ on a manifold $M$: the \emph{compact-open} topology, the
\emph{open} topology and the \emph{fine} topology. The compact-open
topology is coarser than the open topology which in turn is coarser
than the fine topology.
A property $P$ of a metric $g$ is \emph{stable} in a given topology
on $\textrm{Lor}(M)$ if in that topology there is an open
neighborhood of $g$ made of metrics which share property $P$, i.e.
if every sufficiently close metric has the property $P$.

A given property may be stable in some topologies and not in others.
If a topology is coarser than another, it is a stronger requirement
to ask stability in that topology than in the other. For instance,
if a property is stable in the compact-open topology then it is
stable in the open topology which in turn implies the stability in
the fine topology.

Since the properties we want to deal with  are conformally
invariant, it is better to work with topologies on $\textrm{Con}(M)$
instead of $\textrm{Lor}(M)$. A property is \emph{conformally
stable} \cite{beem96} if it holds in an open set of equivalence
classes  on $\textrm{Con}(M)$, but the adjective ``conformally''
will be usually omitted.

In the literature there are two other well known topologies:
Whitney's \emph{fine $C^0$} topology \cite[p. 63]{beem96} and Geroch's
\emph{interval} topology \cite{geroch70}. The first is defined on
$\textrm{Lor}(M)$ and coincides with the open topology, while the
second is defined on $\textrm{Con}(M)$ and it is equivalent to the
quotient of the open topology as proved by Lerner \cite{lerner73}.

If the property $P$ is given by ``$(M,g)$ is causal'' then we shall
speak of ``stable causality" in one topology or the other. If no
mention to the topology is made then it is understood that this
topology is the $C^0$ open topology (or its quotient topology if we
are working on $\textrm{Con}(M)$).

We recall \cite{hawking68,hawking73} that a spacetime $(M,g)$ is
\emph{stably causal} if there exists a Lorentzian metric $\tilde{g}
> g$ such that $(M,\tilde{g})$ is causal. This causality condition
corresponds to  stable causality with respect to the $C^0$ open
topology on $\textrm{Con}(M)$, or equivalently with respect to
Geroch's interval topology \cite{geroch70,hawking71,lerner73}. In
particular stable causality implies  stable causality with respect
to the fine topology. Hawking \cite{hawking71} speculated that these
two notions of  causal stability differ, but actually, as we shall
prove below, they coincide.

We have already recalled that a spacetime $(M,g)$ is \emph{compactly
stably causal} if for every relatively compact open set $V$ there is
a metric $\tilde{g}_V\geq g$ such that $\tilde{g}_V>g$ on $V$,
$\tilde{g}_V=g$ on $M\backslash V$ and $(M,\tilde{g}_V)$ is causal.
Compact stable causality is weaker than stable causality
\cite{minguzzi07d}, thus  the question naturally arises if compact
stable causality can be regarded as a stable causality condition
with respect to a topology finer than the  open topology.  At the
beginning of section  \ref{piero} we argue that no reasonable such
topology exists. Nevertheless, compact stable causality has a
topological origin, indeed it follows from the difference between
interior and sequential interior given Geroch's interval topology on
$\textrm{Con}(M)$ (see Sect. \ref{piero}).



\subsection{Fine topology and stable causality}
We redefine the three topologies introduced by Hawking in his work
in a way which is more convenient for our purposes. In these
definitions there are no requirements on the derivatives of the
metrics, that is, we shall limit ourselves to the $C^0$ topologies.
Unlike Hawking we want to topologize directly $\textrm{Con}(M)$
instead of $\textrm{Lor}(M)$. However, the topologies defined below
are equivalent to the topologies considered by Hawking once one
passes to the quotient space $\textrm{Con}(M)$.

With ``$g$'' we may denote the metric in $\textrm{Lor}(M)$ or the
conformal class of $g$ in $\textrm{Con}(M)$, the meaning being clear
from the context.

\begin{description}
    \item[compact-open topology:]
    If $\underline{g},\overline{g}$,
    are two conformal classes such that $\underline{g}<\overline{g}$
    and $A \subset M$ is an open relatively compact set, the  set $S(A,\underline{g},\overline{g})$
    is defined as the set of all conformal classes $g$ such that
    $\underline{g}<g<\overline{g}$ on $A$. The set of all such $S(A,\underline{g},\overline{g})$ for all
    $A$, $\underline{g}$ and $\overline{g}$, gives a subbasis for the topology,
    i.e. the open sets are the unions of the finite intersections of
    the sets
    $S(A,\underline{g},\overline{g})$. Note that in any open set the  conformal classes are not bounded at infinity.
    \item[open topology:] as above, the subbasis for the topology
    is $S(U,\underline{g},\overline{g})$ but in this case the set $U$ can be any subset
    of $M$, thus without loss of generality we can fix $U=M$. We have $S(M,\underline{g},\overline{g})=\{g\in Con(M): \underline{g}<g<\overline{g}\}$
    thus the topology coincides with {\em Geroch's interval topology} \cite{geroch70}, and the $S$ sets form actually a basis for the topology.
    Note also that in this case the open set places bounds on its elements at infinity.
    \item[fine topology:] let $g$, $\underline{g}$, and $\overline{g}$, be three conformal classes
    such that $\underline{g}<g<\overline{g}$. The set
    $B(g,\underline{g},\overline{g})$ is given by the conformal
    classes $\tilde{g}$ such that $\underline{g}<\tilde{g}<\overline{g}$ and there is an open relatively compact set $A(\tilde{g})$ so that
         $\tilde{g}=g$ outside $A$. The sets $B(g,\underline{g},\overline{g})$ form a subbasis for the
    topology, i.e. the open sets are unions of the finite intersections of these sets.
\end{description}

\begin{remark}
Actually  the sets $B(g,\underline{g},\overline{g})$ used as a
subbasis for the fine topology form a basis of the same topology. In
order to prove this fact we have to show that  the finite
intersections of those sets are an union of $B$ sets, that is, for
every $g$ belonging to the intersection there exists a set $B \ni g$
contained in the intersection.

We prove this fact for an intersection of two sets, the
generalization to finite intersections being straightforward.  Let
$g \in B(g_1,\underline{g}_1,\overline{g}_1) \cap
B(g_2,\underline{g}_2,\overline{g}_2)$, so that
$\underline{g}_1<g<\overline{g}_1$ and
$\underline{g}_2<g<\overline{g}_2$. Since Geroch's intervals form a
base for  Geroch's interval topology there are two metrics
$\underline{g}$, $\overline{g}$, such that
\[\underline{g}_1,\underline{g}_2<\underline{g}<g<\overline{g}<\overline{g}_1,\overline{g}_2.\]

Note that there is an open relatively compact set $A$ such that
$g_1=g_2(=g)$ outside the set $A$. Indeed, a conformal class $g$
belongs to the intersection if and only if
$\underline{g}_1<g<\overline{g}_1$,
$\underline{g}_2<g<\overline{g}_2$, and there exist two open
relatively compact sets $A_1,A_2$ such that
  $g=g_1$ outside
$A_1$ and $g=g_2$ outside $A_2$, thus outside $A=A_1 \cup A_2$ it
must be $g=g_1=g_2$.

Thus $g \in B(g,\underline{g},\overline{g})$ and
$B(g,\underline{g},\overline{g}) \subset
B(g_1,\underline{g}_1,\overline{g}_1) \cap
B(g_2,\underline{g}_2,\overline{g}_2)$.

\end{remark}

\begin{proposition} \label{jsd}
The spacetime  $(M,g)$ is stably causal in the fine topology of
$\textrm{Con}(M)$ if and only if it is stably causal.
\end{proposition}
\begin{proof}
$\Rightarrow$. Let $(M,g)$ be stably causal in the fine topology
then, since the $B$ sets defined above form a basis for the
topology, there exist $\underline{g}$ and $\overline{g}$,
$\underline{g}<g< \overline{g}$, such that
$B(g,\underline{g},\overline{g})$ includes only causal metrics.
Assume that $(M,g)$ is not stably causal then, since stable
causality coincides with stable chronology \cite{minguzzi07},
$(M,\overline{g})$ is not chronological. As a consequence, there
exists a closed $\overline{g}$-timelike curve $\gamma$. Since the
light cones can be narrowed nearby the timelike curve without
spoiling its causal nature, there is a metric $g'$, $g \le g' \leq
\overline{g}$ such that ${g}'<\overline{g}$ on an open relatively
compact set $D$ including $\gamma$, ${g}'=g$ on $M\backslash D$ and
such that $\gamma$ is ${g}'$-causal. Hence ${g}' \in
B(g,\underline{g},\overline{g})$ but $g'$ is not causal, a
contraddiction.

$\Leftarrow$. If $(M,g)$ is not stably causal in the fine topology
then it is not stably causal in the open topology because  the
latter is coarser than the former.
\end{proof}

Hawking \cite{hawking71}  expresses the opinion that stable
causality under the fine topology on $\textrm{Con}(M)$ should be
considerably weaker than stable causality. The previous proposition
shows that this is false and that both topologies lead to stable
causality.

\subsection{Compact stable causality and topology} \label{piero}

Since compact stable causality is weaker than stable causality it
remains the open question of determining whether compact stable
causality can be regarded as a stable causality condition with
respect to a topology finer than the fine topology. We give an
argument which shows that no reasonable topology exists. Suppose
indeed that there exists a topology $\tau$ on $\textrm{Con}(M)$ such
that stable causality with respect to $\tau$ is equivalent to
compact stable causality. Then, given $g \in \textrm{Con}(M)$ such
that $(M,g)$ is compactly stably causal, there exists a $\tau$-open
set $W \subset \textrm{Con}(M)$, $W \ni g$, such that for every $g'
\in W$, $(M,g')$ is causal. But furthermore $(M,g')$ is compactly
stably causal, as $W$ is a neighborhood of causal metrics for $g'$
as well; thus $W$ is made by compactly stably causal metrics.
Consider the example in \cite[Figure 2]{minguzzi07b}, it is a
non-$\overline{A^\infty}$-causal but compactly stably causal
spacetime \cite{minguzzi07d}. In this spacetime, for every open
relatively compact set $V$ containing the displayed point $x$, and
for every metric $g'\geq g$ such that $g'>g$ on $V$ it can be shown
that $(M,g')$ is non compactly stably causal. Hence, every
$\tau$-neighborhood of $g$ does not contain metrics $g'>g$ on the
compact set $\{x\}$. This is clearly an undesirable feature for a
topology as the neighborhoods become too small, in fact so small
that the metrics obtained by slightly perturbing $g$ around $x$
would not belong to a neighborhood of the topology.

Despite the fact that compact stable causality does not come from a
topology, there is a deep and natural topological connection between
compact stable causality and stable causality. Before we explore it,
let us introduce some not well known topological concepts
\cite{snipes72,goreham01}.

Let $(X, \tau)$ be a topological space. Given $A \subset X$, the
\emph{sequential closure} of $A$, written $\textrm{Cl}_s(A)$, is the
union of $A$ and the set of all points in $X$ which are limits of
sequences in $A$. As a consequence, $A\subset
\textrm{Cl}_s(A) \subset \bar{A}$. The topological space is known as
\emph{Fr\'echet-Urysohn} if  $\textrm{Cl}_s(A) = \bar{A}$ for every
$A$. Note that the sequential closure operator is not necessarily an
idempotent operator, i.e. it is not the case that
$\textrm{Cl}_s(\textrm{Cl}_s(A))=\textrm{Cl}_s(A)$ for each subset
$A$ of $X$. The topological spaces that have this property are
called \emph{T-sequential}. The \emph{sequential interior} of $A$,
written $\textrm{Int}_s(A)$, is the set $\textrm{Int}_s(A)=
A\,\backslash \textrm{Cl}_s\!(X\backslash A)$. As a consequence, $
\textrm{Int}(A) \subset \textrm{Int}_s(A) \subset A$. Thus $x \in
\textrm{Int}_s(A)$ if and only if $x \in A$ and there is no sequence
$\{x_n\}$ in $X \backslash A$ such that $\{x_n\}$ is convergent to
$x$. Stated in another way, $x \in \textrm{Int}_s(A)$ if and only if
$x \in A$ and every sequence converging to $x$ is eventually (or
ultimately) in $A$.

Note that a topological space is Fr\'echet-Urysohn if and only if
$\textrm{Int}_s(A)=\textrm{Int}(A)$ for every subset $A$.

The set $A$ is \emph{sequentially closed} if $\textrm{Cl}_s(A)=A$.
Thus $A$ is sequentially closed if  $A$ contains all the points of
$X$ which are limits of sequences in $A$. Since $\textrm{Cl}_s$ is
not idempotent the sequential closure of a set is not necessarily
sequentially closed. Note that a closed set is sequentially closed.

The set $A$ is \emph{sequentially open} if its complement is
sequentially closed. In other words, $A$ is sequentially open if
every sequence converging to a point of  $A$ is ultimately in $A$.
Every open set is sequentially open. Note that if the topological
space is not $T$-sequential, the sequential interior of a set need
not be sequentially open, since the sequential closure need not be
sequentially closed.

Every first countable topological space is Fr\'echet-Urysohn, and in
turn Fr\'echet-Urysohn spaces are $T$-sequential.

We know from Lerner that the interval topology is not  first
countable for non-compact $M$ \cite[Paragraph 2.1]{lerner73}. Actually,
it even fails to be Fr\'echet-Urysohn. As we shall prove below, the
difference between compact stable causality and stable causality
lies in the difference between the sequential interior and the
interior in Geroch's interval topology.

From now on we will consider on $\textrm{Con}(M)$ only Geroch's
interval topology.

The following proposition is known (\cite[Paragraph 2.1]{lerner73},
\cite[p. 448]{geroch70}).

\begin{proposition} \label{intervalConvergence}
Let $M$ be a non-compact Lorentz manifold. The convergence of a
sequence $h_n \to h$ on $\textrm{Con}(M)$ in the interval topology
implies that there exists an open relatively compact set $A \subset
M$ such that for sufficiently large $m$, $h_m=h$ outside $A$.
\end{proposition}

\begin{proof}
Let $p\in M$ and let $B_k$ be the open (relatively compact) balls
centered at $p$ of radius $k$ with respect to a complete riemannian
metric on $M$. If the open relatively compact set in the statement
of the proposition does not exist, there is $n_k>k$ and some $x_k\in
M\backslash B_k$  such that $h_{n_k}(x_k)\ne h(x_k)$.

Note that $x_k \to +\infty$. It is now possible to find metrics
$\underline{h},\overline{h}$, such  that $h \in
(\underline{h},\overline{h})$, and so close to $h$ at the points
$x_k$ that  $h_{n_k}(x_k)\notin
(\underline{h}(x_k),\overline{h}(x_k))$. Thus the interval
$(\underline{h},\overline{h})$ is a neighborhood of $h$ that does
not contain any element of the subsequence $h_{n_k}$, thus $h_n$
does not converge to $h$.
\end{proof}

We denote by $\mathscr{C} \subset \textrm{Con}(M)$  the set of
chronological metrics. It is a well known fact that $\textrm{Con}(M)
\backslash \mathcal{C}$ is open \cite{hawking71} (because a closed
$g$-timelike curve remains timelike in a suitable interval
neighborhood of $g$), hence $\mathcal{C}$ is closed.

The next two theorem clarify the topological relationship between
stable causality and compact stable causality, and in particular the
relationship between compact stable causality and Geroch's interval
topology.
\begin{theorem} \label{Int}
 $g \in \textrm{Int}\mathscr{C}$ if and only if $(M,g)$ is stably
causal.
\end{theorem}
\begin{proof}
$\Rightarrow$. Assume that $g \in \textrm{Int}\mathscr{C}$ so that
there is an interval $(\underline{g},\overline{g})\ni g$ contained
in $\mathscr{C}$ and thus made of chronological metrics. In
particular $(g+\overline{g})/2 \, (>g)$ belongs to the interval and
hence is chronological. Thus $(M,g)$ is stably chronological and
hence stably causal \cite{minguzzi07}.

$\Leftarrow$. Assume that $(M,g)$ is stably causal, then there
exists an open set $(\underline{g},\overline{g}) \ni g$ containing
only causal (and hence chronological) metrics.
\end{proof}

\begin{theorem} \label{Ints}
 $g \in \textrm{Int}_s\mathscr{C}$ if and only if $(M,g)$ is compactly stably causal.
\end{theorem}
\begin{proof}
$\Leftarrow$. Let $(M,g)$ be compactly stably causal and consider a
sequence $g_n \rightarrow g$ in the interval topology, then there
exists a compact set $K$ such that, for sufficiently large $n$,
$g_n=g$ on $M \backslash K$ (it follows from Prop.
\ref{intervalConvergence}). Since $(M,g)$ is compactly stably
causal, there exists a metric $g_K \geq g$ such that $g_K > g$ on
$K$ and $g_K$ is causal and thus every metric narrower than $g_K$ is
also causal. Note that as $K$ is compact, we can find $g'>g$ such
that $g'\le g_K$ on $K$. Since $g_n\to g$, for sufficiently large
$n$, $g_n<g'$ and hence $g_n\le g_K$ on $K$ while $g_n=g$ outside
$K$. For sufficiently large $n$ we have $g_n \leq g_K$, thus for
sufficiently large  $n$ the metrics $g_n$ are all causal and hence
$g_n$ is eventually in $\mathscr{C}$, that is $g \in
\textrm{Int}_s\mathscr{C}$.

$\Rightarrow$. Suppose that $(M,g)$ is non-compactly stably causal,
then there are two cases: either $(M,g)$ is causal or not.

Consider the former case: $(M,g)$ causal. Since $(M,g)$ is
non-compactly stably causal there exists a relatively compact open
set $A$ and a sequence of non-causal metrics $g_n > g$  on $A$,
coinciding with $g$ on $M \backslash A$, such that $g < g_{n+1} <
g_n$ on $A$ and $g_n \rightarrow g$ pointwisely. Since
$\overline{A}$ is compact we have, basically because of Dini's
lemma, $g_n \rightarrow g$ also in the interval topology. Every such
$g_n$ is also non-chronological: indeed $g_{n+1}$ is non-causal and
thus there is a closed $g_{n+1}$-causal curve that necessarily
intersects $A$. Switching to $g_n$, since there is a piece of the
closed curve that is $g_n$-timelike there exists a closed
$g_n$-timelike curve and thus $g_n \in \textrm{Con}(M)\backslash
\mathscr{C}$. We conclude $g \in \textrm{Cl}_s
(\textrm{Con}(M)\backslash \mathscr{C})$, that is $g \notin
\textrm{Int}_s \mathscr{C}$.

Consider the latter case: $(M,g)$ non-causal. Let $A$ be a
relatively compact open set  which contains a closed $g$-causal
curve $\gamma$, and let $g_n\ge g$ be metrics such that
$g<g_{n+1}<g_n$ on $A$ and $g_n=g$ outside $A$ such that $g_n\to g$
pointwisely (and hence also in the interval topology). Clearly
$\gamma$ is $g_n$-timelike so that $g_n \in
\textrm{Con}(M)\backslash \mathscr{C}$. Finally, $g\in \textrm{Cl}_s
(\textrm{Con}(M)\backslash \mathscr{C})$ and hence  $g \notin
\textrm{Int}_s \mathscr{C}$.

\end{proof}

\begin{remark}
Since compact stable causality differs from stable causality
\cite{minguzzi07d}, the previous theorems imply that for generic
$M$, the interval topology on metrics is not  Fr\'echet-Urysohn.
Recall the example in \cite[Figure 2]{minguzzi07b}, already examined
at the beginning of this section, and consider a sequence $g_n
\rightarrow g$ in the interval topology, such that $g \leq g_{n+1}
\leq g_n$ and the strict inequality holds on an open relatively
compact set $A$ containing the displayed point $x$. Even though
$(M,g)$ is compactly stably causal it is easy to check that, for
every $n$, $(M,g_n)$ is non-compactly stably causal, that is $g_n
\notin \textrm{Int}_s\mathscr{C}$. Hence $g_n$ is not eventually in
$\textrm{Int}_s\mathscr{C}$ and thus $g \notin
\textrm{Int}_s\textrm{Int}_s\mathscr{C}$. As a consequence,
$\textrm{Int}_s\textrm{Int}_s\mathscr{C} \neq
\textrm{Int}_s\mathscr{C}$ and the interval topology is not
$T$-sequential. This argument holds for the particular manifold
given by the spacetime of  \cite[Figure 2]{minguzzi07b},
nevertheless the conclusion holds for general $M$ as the following
proposition shows.
\end{remark}

\begin{proposition}
Let $M$ be a non-compact manifold. The interval topology on
$\textrm{Con}(M)$ is not $T$-sequential.
\end{proposition}

\begin{proof}
Let $B \subset M$ be a relatively compact open set and let $w_{n},y
\in \textrm{Con}(M)$ be such that  $y\!< \! w_{n+1}\! <\! w_n $ on
$B$ and $w_n = y$ on $M \backslash B$, and $w_{n}
\xrightarrow{n\to+\infty} y$ pointwisely and thus, being $\bar{B}$
compact, in the interval topology. Let $B_{n}\subset M$ be a
sequence of disjoint relatively compact open sets such that every
compact set of $M$ contains at most a finite number of the $B_n$'s
(i.e. the sets $B_n$ go to infinity). We can assume $B_n\cap
B=\emptyset$. For every $n$, consider a sequence $s_n^{m} \subset
\textrm{Con}(M)$  such that $w_n\!<\!s_n^{m+1}\!<\!s_n^{m}$ on
$B_n$, and $w_n \!=\! s_n^{m}$ on $M \backslash B_n$ and $s_n^{m}
\xrightarrow{m\to +\infty} w_n$, pointwisely and thus, being
$\bar{B}_n$  compact, in the interval topology. Consider the set
$S=\{s_n^m: n,m \in \varmathbb{N}\}$, by construction $w_n \in
\textrm{Cl}_s S$ and $y \in \textrm{Cl}_s (\textrm{Cl}_s S)$. We are
going to show that $y \notin \textrm{Cl}_s S$, because there is no
sequence $s_{n(k)}^{m(k)} \rightarrow y$. Assume such sequence
exists. From proposition \ref{intervalConvergence} an open
relatively compact set $A$ would exist such that $s_{n(k)}^{m(k)} =
y$ on $M \backslash A$ for sufficiently large $k$. Since on
$B_{n(k)}$ we have $s_{n(k)}^{m(k)}> w_{n(k)}=y$ it must be for
sufficienlty large $k$, $ B_{n(k)} \subset A$.  As $A$ is a
relatively compact set there is some $\overline{n} \in
\varmathbb{N}$ such that $n(k) < \overline{n}$ and hence on $B$ for
sufficiently large $k$, $s_{n(k)}^{m(k)}=w_{n(k)} > w_{\overline{n}}
> y$, hence since $\bar{n}$ does not depend on $k$ there is no
convergence to $y$. Thus $\textrm{Cl}_s (\textrm{Cl}_s S) \neq
\textrm{Cl}_s S$.
\end{proof}

In \cite{hawking71}, Hawking conjectures that it is \emph{generic} for a
metric satisfying ordinary causality  to satisfy stable causality,
i.e. that stably causal metrics are dense in the causal metrics. The
following proposition gives a simple proof that the conjecture is
true.

\begin{proposition} \label{cfa}
The set of stably causal metrics of $\textrm{Con}(M)$, $\textrm{Int}
\mathscr{C}$, is dense in the set of chronological metrics i.e.
$\overline{\textrm{Int} \mathscr{C}}=\mathscr{C}$.
\end{proposition}
\begin{proof}
We have to prove that $\overline{\textrm{Int}
\mathscr{C}}=\mathscr{C}$. The set $\mathscr{C}$ is closed, since
the set of non-chronological metrics $M \backslash \mathscr{C}$ is
open \cite{hawking71}. Thus $\textrm{Int} \mathscr{C} \subset
\mathscr{C} \Rightarrow \overline{\textrm{Int} \mathscr{C}} \subset
\overline{\mathscr{C}} = \mathscr{C}$. It remains to show that
$\mathscr{C} \subset \overline{\textrm{Int} \mathscr{C}}$. Suppose
that there exists $g \in \mathscr{C}\backslash
\overline{\textrm{Int} \mathscr{C}}$, it means that $g$ is
chronological but there exists an open neighborhood
$(\underline{g},\overline{g})$ of $g$ that does not contains any
point of $\textrm{Int} \mathscr{C}$, that is any stably causal
metric. This is false, given that $\underline{g}\!<\!g$ and
$g'\!=\!(\underline{g}+g)/2$ is such that
$\underline{g}<g'<g<\overline{g}$. Hence $g'$ is stably
chronological thus stably causal \cite{minguzzi07}, a contradiction.
\end{proof}

\section{Conclusions}

In this work we have investigated the stability of causality under
perturbations of the metric at infinity or in finite spacetime
regions. We have shown that for non-imprisoning spacetimes stable
causality is equivalent to the possibility of widening the cones
outside any chosen compact set without spoiling causality. This
result has been obtained by using the recently proved equivalence
between stable causality and $K$-causality.

On a dual direction we have considered what happens widening the
light cones inside the compact sets. If the spacetime is compactly
stably causal this operation can be done without spoiling causality.
We have shown  that compact stable causality corresponds to the
antisymmetry condition of a transitive (but in general non closed)
relation $J^{+}_{CS}$ that we have explicitly constructed. This
result is analogous to the one which states that stable causality is
equivalent to the antisymmetry of the Seifert relation $J^{+}_S$.

We have argued that compact stable causality can not be obtained as
a causal stability condition with respect to a suitable topology on
metrics. Nevertheless, compact stable causality is nicely related to
the Geroch's interval topology on $\textrm{Con}(M)$. Indeed, we
proved that the compactly stably causal metrics are exactly those in
the {\em sequential} interior of the set of chronological metrics,
while the stably causal metrics are those in the usual interior. The
difference between the two interior concepts arises because the
Geroch's interval topology is not Fr\'echet-Urysohn and in fact we
have shown that it is not even $T$-sequential.

Other results include the proof that the causal stability condition
with respect to the $C^0$ fine topology leads to the usual notion of
stable causality (Prop. \ref{jsd}), and the proof that the stably
causal metrics are dense in the set of chronological metrics (Prop.
\ref{cfa}).

\section*{Acknowledgments}
The authors thank Steven Harris and Robert Low for reading the
manuscript and giving many useful suggestions. This work has been
partially supported by GNFM of INDAM.


\end{document}